# The enigmatic abundance of atomic hydrogen in Saturn's upper atmosphere

Short title: The Saturn Lyman-α Bulge

Lotfi Ben-Jaffel[1,2], Julie Moses[3], Robert A. West[4], M-K. Aye[5], Eric T. Bradley[6], John T. Clarke[7], Jay B. Holberg[2], Gilda E. Ballester[2]

---

[1] Sorbonne Universités, UPMC Univ Paris 6 & CNRS, UMR 7095, Institut d'Astrophysique de Paris, 98 bis bd Arago, F-75014 Paris, France; bjaffel@iap.fr
[2] University of Arizona, Lunar & Planetary Laboratory, Tucson, AZ, USA
[3] Space Science Institute, Boulder, CO, USA
[4] Jet Propulsion Laboratory, Pasadena, CA, USA
[5] LASP, University of Colorado, Boulder, CO, USA
[6] University of Central Florida, Orlando, FL, USA
[7] Boston University, Boston, MA, USA

Abstract

A planet's Lyman-α (Lyα) emission is sensitive to its thermospheric structure. Here, we report joint Hubble Space Telescope (HST) and Cassini cross-calibration observations of the Saturn Lyα emission made two weeks before the Cassini grand finale. To investigate the long-term Saturn Lyα airglow observed by different ultraviolet instruments, we cross-correlate their calibration, finding that while the official Cassini/UVIS sensitivity should be lowered by ~75%, the Voyager 1/UVS sensitivities should be enhanced by ~20% at the Lyα channels. This comparison also allowed us to discover a permanent feature of the Saturn disk Lyα brightness that appears at all longitudes as a brightness excess (Lyα bulge) of ~30% (~12σ) extending over the latitude range ~5-35N compared to the regions at equator and ~60N. This feature is confirmed by three distinct instruments between 1980 & 2017 in the Saturn north hemisphere. To analyze the Lyα observations, we use a radiation transfer (RT) model of resonant scattering of solar and interplanetary Lyα photons, and a latitude-dependent photochemistry model of the upper atmosphere constrained by occultation and remote-sensing observations. For each latitude, we show that the Lyα observations are sensitive to the temperature profile in the upper stratosphere and lower thermosphere, thus providing useful information in a region of the atmosphere that is difficult to probe by other means. In the Saturn Lyα bulge region, at latitudes between ~5 to ~35°, the observed brightening and line broadening support seasonal effects, variation of the temperature vertical profile, and potential superthermal atoms that require confirmation.

1. Introduction

The global energy balance in the Earth's thermosphere is predominantly a product of solar heating. In contrast, the thermospheres of all the outer planets are several times hotter (500-1100 K) than they would be merely from heating by solar radiation (< 200 K), leading to a very intriguing energy crisis. Analysis of many Cassini UVIS occultation observations of Saturn has confirmed a high temperature for its thermosphere, with a net temperature decrease from polar latitudes (~590 K) to the equator (~370 K) (Koskinen et al. 2013, 2015; Brown et al. 2020).

Besides the thermal structure diagnostic, atomic H, which is a significant component and sensitive tracer of the upper atmosphere of Saturn, can strongly constrain the composition and energy budget of its thermosphere. However, the H content remains unknown and is poorly constrained by occultation observations (Koskinen et al. 2020). On the other hand, the use of planetary airglow observations should be straightforward for deriving atmospheric properties and composition. For example, the shape of the Lyα line profile bears key information that has been successfully used in the past to study the atmospheres of all the outer planets as well as their satellites to help determine composition, thermal structure, velocity distributions, and non-thermal processes operating in the upper atmospheric layers (Clarke et al., 1991; Ben-Jaffel et al., 1995; Ben-Jaffel et al. 2007; Chaufray et al. 2010; Strobel et al. 2019).

For instance, Lyα observation of the disk of Saturn started as early as 1976 during minimum solar activity with both a rocket and the Copernicus satellite (Weiser et al. 1977; Barker et al. 1980). In 1980 and 1981, Voyager encounters with the Saturn system allowed a series of observations of the Lyα airglow of the planet during solar maximum activity (Broadfoot et al. 1981, Sandel et al. 1982). In parallel, IUE observations began in 1980 remote observation of the planet at Lyα with a monitoring that lasted until the end of 1990 (Clarke et al. 1981; McGrath & Clarke 1992). One of the first high-resolution observations of the Saturn Lyα non-auroral emission line was obtained with HST/GHRS in 1996. When the Cassini mission began in 2004, a large set of observations of the planet's Lyα brightness was obtained until the end of the mission in September 2017. Finally, during a Cassini/HST joint campaign in 2017, Saturn's airglow was recorded simultaneously by both Cassini/UVIS and HST/STIS Echelle spectrometers, offering a rare opportunity to assess the UVIS calibration using the well-accepted HST calibration as a reference (Bohlin et al. 2019).

One of the complexities of remote observation of Saturn Lyα dayglow is the absorption by the intervening interplanetary medium (IPM) hydrogen between Saturn and Earth. As discussed in the literature, the IPM absorption is Doppler shifted with a spectral position and strength that depend on the heliospheric position of the target with respect to the ISM flow. For example, during the HST/GHRS observation in 1996 (Table 2), the IPM absorption was near the Saturn line center, thus necessitating a model of the IPM extinction in order to deduce the undisturbed Lyα brightness from the one measured by HST/GHRS (e.g., Table 2). As we show below, the new dataset obtained by HST/STIS in 2017 does not require such modeling and deduction to obtain the undisturbed Lyα brightness.

In addition to the IPM extinction, degeneracy in theoretical modeling of the airglow and uncertainty related to instrument calibration (see appendix 1-3) and solar flux variability have made the technique questionable and should therefore be carefully addressed (Ben-Jaffel & Holberg, 2016 ; Strobel et al., 2019). For instance, long-term monitoring of the Lyα airglow of Saturn with many space missions has clearly shown that the bright Lyα emission exhibits a strong correlation with the solar cycle, supporting resonant backscattering of the solar and IPM Lyα emission lines as the dominant sources (McGrath & Clarke 1992; Ben-Jaffel et al. 1995). However, contradictory conclusions resulted from the existing data analysis due to inter-calibration issues between instruments (Ben-jaffel et al. 1995; Koskinen et al. 2020). Using Voyager UVS observations,

Ben-Jaffel et al. (1995) stressed the importance of the IPM Lyα emission reflected by the upper atmosphere of Saturn and concluded that there is a need for an enhanced H content of the upper atmosphere ([H]~9x10$^{16}$ cm$^{-2}$) in order to fit the disk Lyα brightness ~3.3 KR observed by Voyager 1 UVS during solar maximum activity in 1980. In contrast, Koskinen et al. (2020) estimated that the reflected IPM Lyα emission is negligible, and their prediction for the H content ([H]~3x10$^{16}$ cm$^{-2}$) is enough to reproduce Saturn's brightness observed by Cassini/UVIS in 2007 during solar minimum activity. Following Quemerais et al. (2013), they concluded that the Voyager 1 & 2 UVS calibration should then be strongly corrected.

It is important to stress that while all studies described above are derived on the basis of the Lyα line integrated brightness, they are missing the key spectral information of the line profile. Besides calibration issues related to each instrument, the broad Lyα line profile observations by HST/GHRS in 1996 question the conclusions obtained only on the basis of integrated brightness (Ben-Jaffel & Holberg, 2016). In addition, we note that past conclusions regarding UVIS & UVS calibration are not adequate because of the absence of a reference instrument observing the genuine Saturn Lyα brightness in the same conditions. The 2017 HST/Cassini campaign reported here addresses that issue.

In the following, we describe the new data sets obtained (Section 2). In Section 3, starting from existing official pipelines, we re-assess the calibration of UVIS and STIS around the Lyα spectral band. Independently of any theoretical modeling, we also cross-correlate the calibration of UVIS and STIS with the International Ultraviolet Explorer (IUE) spectrometer, the Hubble Space Telescope/Goddard High Resolution Spectrometer (GHRS), and Voyager 1 & 2 UV Spectrometers, all of which observed Saturn Lyα airglow over the 1980-2017 period. In Section 4, we use an adding-doubling radiative transfer model to analyze Saturn high-resolution Lyα line profiles observed by HST/STIS and the disk brightness distribution observed by several UV instruments. Finally, we discuss our finding within the global context of Saturn's complex upper atmosphere, focusing on new modeling efforts that will be required in the future.

## 2. Observations

On August 26, 2017, HST/STIS performed high-resolution (Δλ ~ 0.006 nm) spectro-imaging of Saturn's dayglow using the Echelle E140H grating with the long 52"x0.5" slit (HST/GO 14931) simultaneously with Cassini/UVIS low-resolution (Δλ ~ 0.48 nm) measurement of the night-side brightness. This was the first step of the HST/Cassini campaign that was intended to disentangle the IPM source contribution to the planetary Lyα emission. In the second step on September 2, 2017, HST/STIS obtained another high-resolution spectral imaging of the planetary dayglow with a different geometry of observation simultaneously with a Cassini/UVIS scan of the planetary limb. On both dates, HST/STIS also performed NUV imaging of Saturn using a near UV filter (F25srf2) that helped accurately capture the geometry of observation of the oblate shape of the planet. This important step allowed accurate definition the light scattering conditions at the different locations observed over the planetary disk. Details about all observations obtained during the campaign for both STIS and UVIS are listed in Table 1.

Here, it is important to stress that using HST/STIS E140H Echelle grating is crucial to achieve the high resolution measurement required by the RT analysis of the planetary Lyα line profile. When associated with a narrow and short slit, the STIS Echelle mode works well and has been used in the past to investigate auroral emissions from other planets (Chaufray et al., 2010). However, when associated with the long planetary slit, the Echelle mode is allowed for observations but not fully supported in the STSCI Calstis calibration pipeline. Indeed, when using this mode for extended sources, it is difficult to handle overlap between Echelle orders at different positions along the long slit spatial direction, thereby making the extraction of the target signal very difficult. However, as shown in Vincent et al. (2011 & 2014), there is no issue for cold planetary atmospheres or sky background FUV emission where only the Lyα line is dominant. In addition, even if a few Earth geocorona airglow lines are present, they are several orders of magnitude fainter that the Lyα signal (Vincent et al. 2014). In that frame, as far as we avoid the very bright auroral region, the E140H long-slit Echelle mode works well for the Saturn Lyα study. In practice, we have verified that no inter-order contamination is affecting our data, which is consistent with long-slit mapping of the Saturn disk outside the polar region. If any faint auroral contamination appears, it must be below the 1 sigma statistical noise.

In that frame, we calibrated the HST/STIS E140H long-slit Echelle data using the procedure described in detail in appendix 2 (see also Vincent et al. 2011, 2014). In summary, for each exposure, starting from the flat-field file provided by STSCI, the pipeline consists in correcting the geometry distortion of the spectral image, subtracting the so-called detector FUV glow that is monitored by STSCI over time, and using a new technique for removing the contamination from Earth's geocorona and the sky background Lyα emissions (see Appendix 2 for more details). To obtain the sky background emission line, we have performed a STIS/E140H spectral imaging of the sky far from the planetary disk during one HST orbit (e.g., Table 1). For reference, the flux calibration is performed using the standard STIS/E140H calibration for extended targets[8] (e.g., Appendix 2).

---

[8] https://hst-docs.stsci.edu/stisdhb/chapter-5-stis-data-analysis/5-4-working-with-spectral-images

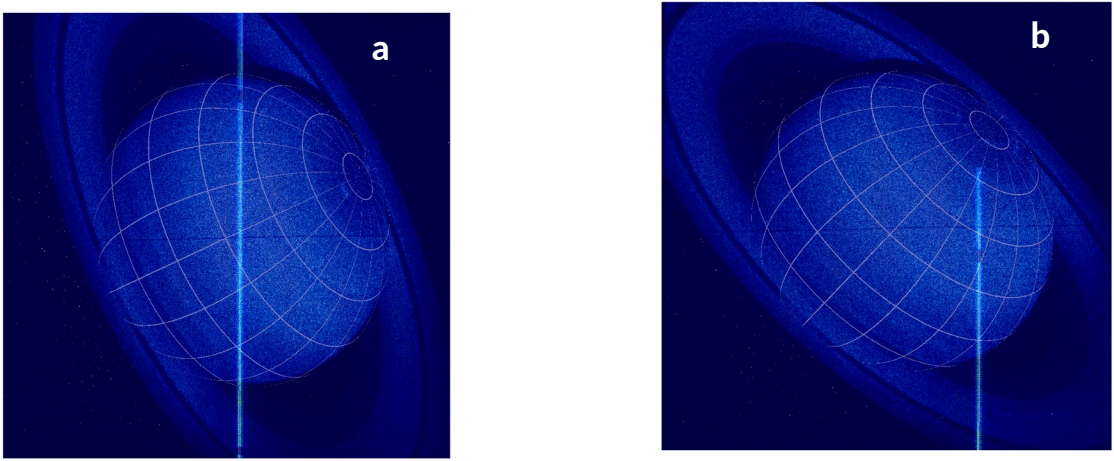

**Figure 1:** NUV image of Saturn showing the geometry of the HST/STIS observations. The STIS 52"x0.5" long slit is shown, particularly the fiducial bars that block light from two regions along the slit and appear as gaps. For August 26, 2017, the slit center was at CML, while for September 02, the slit was shifted from CML to avoid the auroral region. At both dates, only the north pole is visible. **(a)** Sketch for the August 26, 2017 observations. **(b)** Same, but for the September 2, 2017 observations.

Remarkably, that during the HST/Cassini campaign in August/September 2017, the spectral position of the IPM absorption (~$10^4$ K) was strongly blue-shifted (~23 km/s) relative to the Saturn reference frame. Thus, an important consequence is that the red-side half of the Saturn Lyα line profile remains unaffected by the IPM absorption. In the worst-case scenario of a dense IPM, this absorption of the red wing should not exceed ~3%. This means that the HST/STIS observation provides the genuine undisturbed half line of the planetary emission (or at least a minimum value) that can be directly compared to UVIS in situ observations.

To improve the signal to noise for each date of observation, we merged all available HST exposures (e.g., Table 1) and binned by 50 pixels (~ 1.45") along the slit spatial direction.

**Table 1:** HST/STIS[9] & Cassini/UVIS observations obtained during the HST/Cassini 2017 campaign. We also use other archive data of different UV instruments that we describe in Section 3.

| Instrument/Grating | Mode | Data set | Obs. date | Obs. time | Expo. Time (s) | Description |
|---|---|---|---|---|---|---|
| HST/E140H | 52x0.5 | ODFBA1010 | 2017-08-26 | 02:40:24 | 1861.148 | Saturn disk |
| HST/E140H | 52x0.5 | ODFBA1020 | 2017-08-26 | 03:57:17 | 2941.195 | Saturn disk |
| HST/E140H | 52x0.5 | ODFBA1030 | 2017-08-26 | 05:33:21 | 2898.193 | Saturn disk |
| HST/G140M | 52x0.1 | ODFBA1040 | 2017-08-26 | 07:07:59 | 597.001 | Saturn disk |
| HST/MIRFUV | F25SRF2 | ODFBA1D8Q | 2017-08-26 | 07:41:33 | 907.200 | NUV disk |
| HST/E140H | 52x0.5 | ODFB1A010 | 2017-08-26 | 08:47:54 | 2483.184 | Sky Back. |
| HST/E140H | 52x0.5 | ODFBA2010 | 2017-09-01 | 23:57:21 | 1861.190 | Saturn disk |
| HST/E140H | 52x0.5 | ODFBA2020 | 2017-09-02 | 01:13:42 | 2941.199 | Saturn disk |
| HST/E140H | 52x0.5 | ODFBA2030 | 2017-09-02 | 02:49:47 | 2898.19 | Saturn disk |
| HST/G140M | 52x0.1 | ODFBA2040 | 2017-09-02 | 04:24:25 | 547.014 | Saturn disk |
| HST/MIRFUV | F25SRF2 | ODFBB2DIQ | 2017-09-02 | 05:00:02 | 784.200 | NUV disk |
| HST/E140H | 52x0.5 | ODFB2A010 | 2017-09-02 | 06:43:37 | 2483.193 | Sky Back. |
| Cassini/UVIS | LR | 290SA-EQUAMAP001_VIMS | 2017-08-26 | 10:39:36 | 12240 | Saturn Night |
| Cassini/UVIS | LR | 290SW-IPHSURVEY | 2017-08-27 | 13:55:33 | 28712.5 | Sky Back. |
| Cassini/UVIS | LR | 291SW-IPHSURVEY | 2017-08-31 | 03:02:33 | 25575 | Sky Back. |
| Cassini/UVIS | LR | 291SA-LIMBINT001_PRIME | 2017-09-02 | 02:08:33 | 20160 | Saturn limb |
| Cassini/UVIS | LR | FUV1999-231 | 1999-08-19 | 01:30:40 | 64800 | Sky Back |
| Cassini/UVIS | LR | FUV2007-246 | 2007-09-03 | 01:12:28 | 12880 | Saturn Disk |
| Cassini/UVIS | LR | FUV2013-133 | 2013-05-13 | 00:11:59 | 28800 | Saturn Disk |
| Cassini/UVIS | LR | FUV2013-154 | 2013-06-03 | 15:42:03 | 20880 | Saturn Disk |
| Cassini/UVIS | LR | FUV2014-095 | 2014-04-05 | 14:25:03 | 15600 | Saturn Disk |
| Cassini/UVIS | LR | FUV2014-100 | 2014-04-10 | 14:00:19 | 38400 | Saturn Night |
| Cassini/UVIS | LR | FUV2017-016 | 2017-01-16 | 10:12:33 | 2880 | Saturn Disk |
| Cassini/UVIS | LR | FUV2017-018 | 2017-01-18 | 00:05:46 | 4800 | Saturn Night |

---

[9] HST data accessible via https://doi.org/10.17909/cafj-3r46/

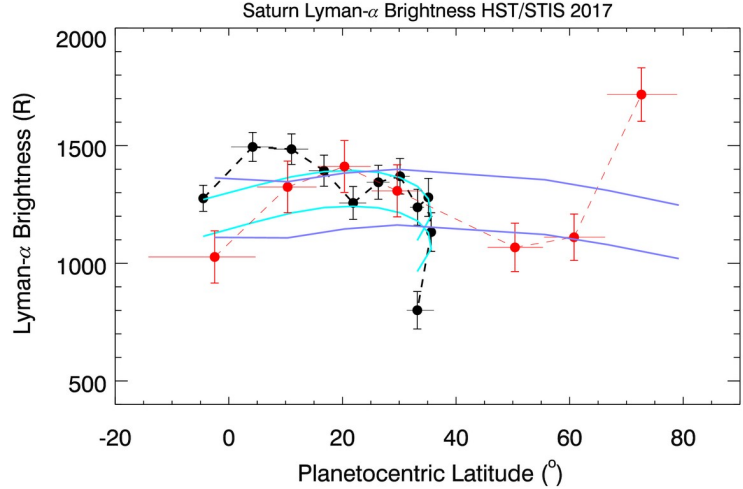

**Figure 2:** Saturn dayside Lyα total brightness versus planetocentric latitude, observed by HST/STIS during the HST/Cassini 2017 campaign. All brightnesses are scaled to the September 2, 2017 solar flux condition. Data averaged over 50 pixels along the HST/STIS long slit (1 pixel~ 0.029" along the spatial direction). We show August 26, 2017 brightness (black), September 2, 2017 (red), and RT model brightness for 2x and 3x the reference HI content (aqua for August 26 RT model and blue for September 02). We provide the reference HI content versus latitude in Section 4.

For HST observations, in order to define the light-scattering geometry over the planetary disk that is required for any radiation transfer modeling, we first fit the oblate shape of the planet using the NUV image and derive the scattering angles (incident solar angle and emission angle) over the planetary disk, particularly along the projection of the STIS long slit over the planetary disk (e.g., Figure 1).

For Cassini/UVIS observations, the scattering geometry is provided by the Cubegenerator, which uses SPICE, a NASA/JPL toolkit software[10].

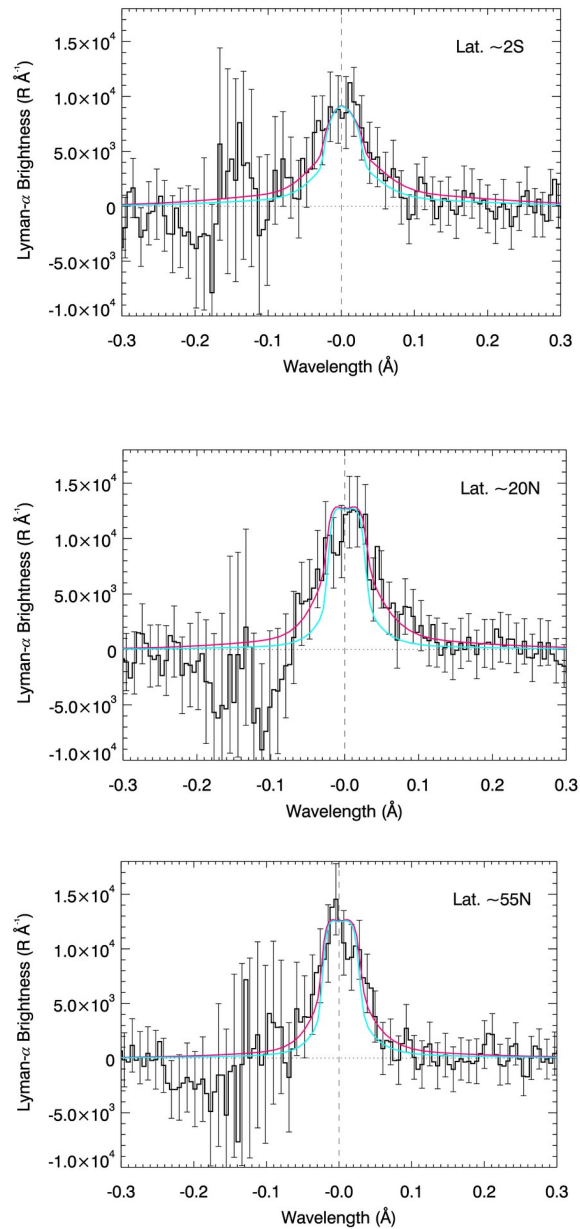

**Figure 3:** Saturn Lyα line profiles observed by HST/STIS (E140H echelle grating and the long slit 52"x0.5"). Only the red-side half of lines is used in

---

[10] https://naif.jpl.nasa.gov/naif/toolkit.html

the brightness diagnostic. Line profiles are shown at specific latitudes indicated in each panel (averaged over the latitude range highlighted in Fig. 2 ). We show RT line profiles models using a reference photochemistry model (aqua) and a best fit (red) corresponding to 2 to 3 times enhanced HI column atmospheric models. Only the red-side half of the line is used in our analysis. Top: Line profile in the latitude range [-13°,5°] when the region was near the planetary limb, a position that explains the line broadening that the RT model reproduces well for a HI column 2x the reference model. Middle: Line profile in the latitude range [16°,25°]. For this position, the RT model requires a HI column 3x the reference model. Bottom: Line profile in the latitude range [50°,60°]. For this position, the RT model requires a HI column 2x the reference model.

The resulting Lyα brightness levels respectively for August 26 and September 2, 2017 are comparable after correcting for the solar flux difference between the two dates (e.g., Figure 2). First, we remark that the September 2, 2017 data set is much closer to a latitudinal distribution (along meridian, e.g., Figure 1). In contrast, the August 26 observations better show the effect of the incident/emission scattering angles. A dip in the brightness distribution of August 26 appears around latitude ~20N but cannot be confirmed because of the noise level. For reference, the solar flux on September 2, 2017 is close to a minimum of solar activity and will be used in the following as our reference level. The change is remarkable in the line profile versus latitude (and also scattering geometry) that we show in detail using the September 2, 2017 observations that cover an extended latitudinal range from the equator up to the north pole (e.g., Figure 3). The STIS Lyα brightnesses will be used as our reference for the cross-correlation with past UV instruments that we investigate in the following section.

### 3. Cross-calibration of UV instruments (1980-2017):

For reference, we describe the STIS E140H long-slit mode calibration steps in Appendix 2, and those of UVIS in Appendix 3.

In the following, we cross-calibrate STIS, UVIS, IUE, Voyager 1 &2 UVS, and HST/GHRS instruments, using Lyα archive observations of Saturn obtained between 1980 and 2017 which cover nearly four consecutive solar cycles. To achieve our goal, we only assume that the Saturn Lyα brightness is resonance back-scattering of solar and interplanetary light sources by its upper atmosphere. Indeed, as discussed in the introduction section, there is strong evidence that this should be the case (McGrath & Clarke, 1992; Ben-Jaffel et al., 1995), yet we have no idea how it is achieved, and we make no specific assumption about the atmospheric composition. To implement the comparison while avoiding theoretical model uncertainty when possible, we use four key parameters to describe the scattering processes that produce the Lyα emission of the planet, namely: phase angle between incident light and observer line of sight, incident angle of the incoming light source, emission angle of the emitted light measured from the local normal to the surface, and finally, the planetocentric latitude location. As shown below (Figure 7), we found no apparent dependency of the Lyα brightness of Saturn to the planetary longitude (Yelle et al., 1986). Finally, we must point out that, when needed, we do include a slight correction (~10%) for the change of the atmospheric reflectivity between minimum and maximum solar activity corresponding to the change in the atmospheric composition (Figure 8). This correction is included for consistency but has no incidence on the conclusion on the calibration correction derived for the different instruments. Because measurements for our reference HST instruments were made during solar minimum activity, if one does not include the evolution of the atmospheric reflectivity versus solar activity, the derived calibration correction factor should be considered as a minimum value. In that frame, our cross-calibration correction factors (minimum) are model-free, particularly for the comparison between STIS and UVIS.

One key input also required to achieve the cross-calibration is an accurate estimation of the solar flux at the Saturn orbit for each date of observation. Usually, the flux at the solar Lyα line center is taken as a reference in the literature. For consistency, we use the Lisird Colorado database that provides the solar Lyα line profile at 1AU, a combination of measurements from multiple instruments and models to estimate the full disk integrated solar Lyα line profile over time (Machol et al., 2019). In addition, we take into account the change in the solar disk irradiance due to the solar rotation between the date of Saturn observation and the Lisird daily prediction/measurement. In a second step, we propagate the 1AU line profile to the orbital position of Saturn (Table 2), taking into account the absorption by the interplanetary hydrogen (IPH) between the Sun and the planet (Wu & Judge, 1979). The IPH absorption depends on the Saturn orbital position in the heliosphere that we describe by the angle between the Sun-Saturn and the interstellar upwind directions and the distance from the Sun (e.g., Table 2). As shown in Figure 4, the imprint of the IPH absorption is not negligible, showing an asymmetric line profile *F(x)* (where *x* is wavelength relative to the line center), which affects the flux level near the line center. Recall that the flux of interest for radiation transfer calculations of the planetary brightness is the mean (F(x)+F(-x))/2 (e.g., Figure 4, magenta curve). For example, for the August 26, 2017 observations, the average flux around the line center (+/- 0.15A from line center) is ~12% smaller when the IPH absorption is included.

For the different observations included here, the line center (LC) average fluxes are shown in Table 2 and will be used for scaling the Saturn Lyα brightness obtained at different dates between 1980 and 2017.

Fortunately, during the lifetime of the Cassini mission, a rich database was collected between 2006 and 2017, sampling extended scattering properties over time. Therefore, starting from the scattering properties obtained during the HST/STIS observations (Section 2), we used the OPUS powerful engine[11] to search in the PDS archive for UVIS observations obtained in the same scattering properties (within ~1° angular error bar) as for each of the two HST/STIS datasets separately (Aug 26 and Sep 02, 2017).

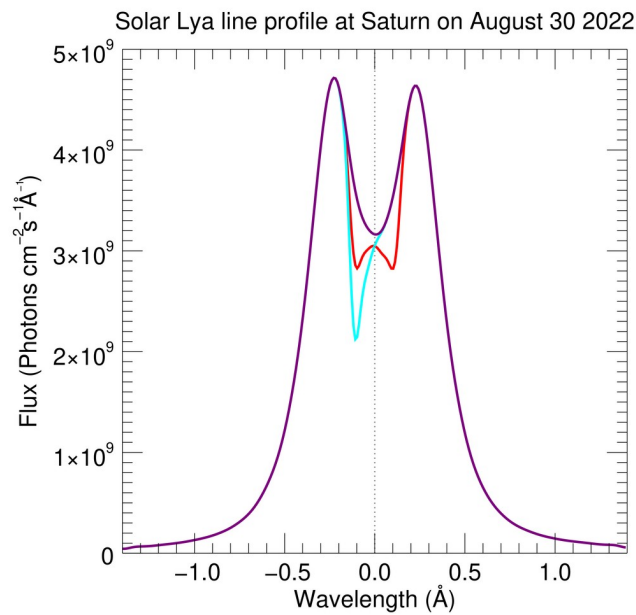

**Figure 4**: Lisird Lyα line profile at Saturn orbital position on August 30, 2017 corresponding to STIS observation on August 26, 2017 (scaled down by the square of the Saturn orbital distance (10.06 AU), and taking into account the Sun rotation. Without the imprint of the interplanetary atomic hydrogen absorption between the Sun and Saturn (magenta). Same but now including

---

[11] https://opus.pds-rings.seti.org/

the IPH absorption that produces an asymmetric line profile (aqua). Line profile as seen in rest frame of the atmosphere *(F(x)+F(-x))/2* that must be used for RT modelling or for scaling solar fluxes between different dates (red).

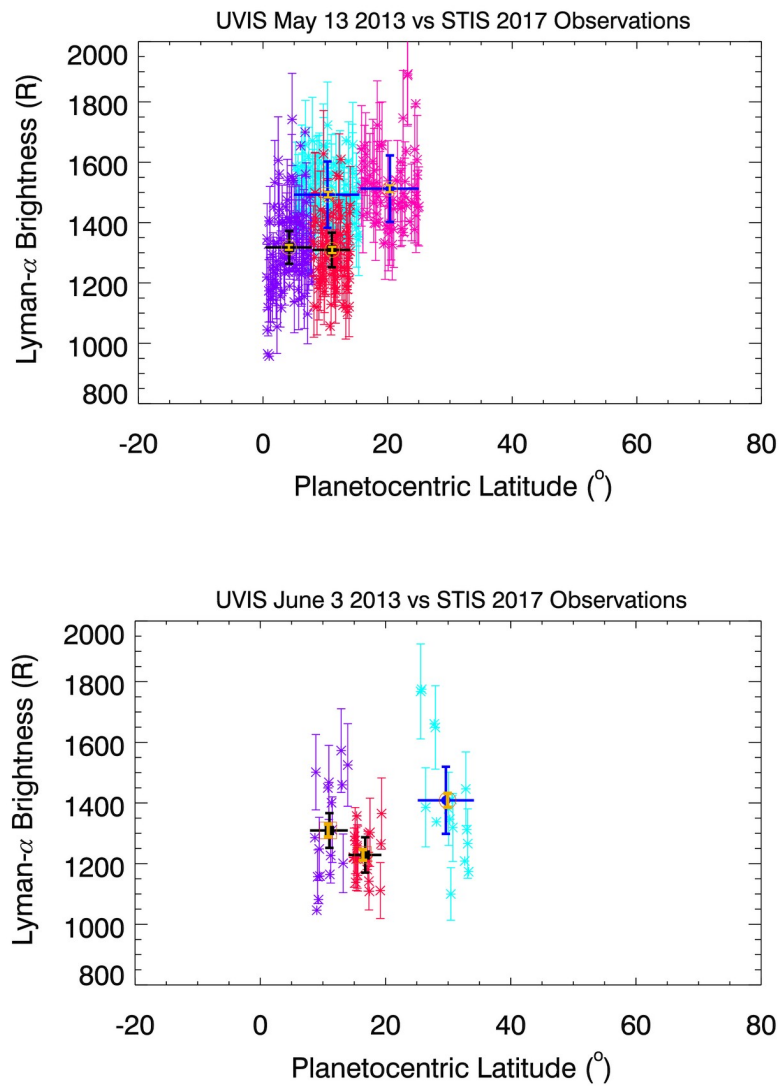

**Figure 5** Comparison between UVIS (scaled by a factor listed below) and STIS observations of Saturn Lyα brightness. Clouds of UVIS data points (colored) that fit the same range of scattering properties as the STIS data points (averaged

over the same latitude range shown by the STIS horizontal error bars). For each date of HST observations (black for Aug 26 and blue for Sep 2 shown in Figure 2), we could find at most two STIS data points for which UVIS data share the same scattering properties. Average of UVIS data points are shown with attached error bar (gold). (Top) UVIS May 13, 2013 data points and related errors scaled (Table 1): purple and red for August 26, 2017 scattering conditions, and aqua and pink for September 2, 2017 scattering conditions. Scaling factors are listed in Table 2, leading to an average correction factor of 1.73±0.05 on UVIS brightness in order to match STIS brightness. (Bottom) Same but for UVIS June 3, 2013 (Table 1). The average correction factor is 1.75±0.06.

We also required that UVIS data be in non-binned mode (1024 spectral pixels and 64 spatial pixels along the slit, Esposito et al. 2004) to avoid the effect of unexpected "evil" pixels. We could find two datasets obtained on May 13 and June 3, 2013 that fulfill those conditions. In both cases, using either August or September STIS data sets, we could match STIS and UVIS brightness levels only if UVIS brightnesses are scaled up by an average factor ~1.75 (e.g. Figure 5), after correcting for the solar flux ratio between the different dates (e.g., Table 2). This scaling is equivalent to reducing the UVIS sensitivity by the same factor.

In the following step, we compare STIS and Voyager V1 & V2 observations, after correction for the solar flux ratio between the different dates (Table 2). We find that while V1 brightnesses need to be scaled down by a factor ~0.8, V2 UVS calibration is compatible with STIS (a correction factor ~0.95 on V2).

For GHRS observations, we found a UVIS dataset obtained on September 3, 2007 that has comparable scattering properties and comparable solar flux (Tables 1 & 2). Again, after correcting for the solar flux ratio between the two dates, an extra correction ~1.78 on the UVIS brightness is required to match the 1996 GHRS measurement (e.g., Table 2).

To compare Voyager V1 to UVIS, we also used OPUS/PDS to find a UVIS dataset obtained on April 5, 2014 that has similar scattering conditions and comparable maximum solar activity conditions. After correcting for the solar flux ratio between the dates, we could fit V1 and UVIS distributions using correction factors for the calibration comparable to those derived from the cross-calibration between STIS & V1 (~0.8) and separately from STIS & UVIS (1.74) (e.g., Table 2).

From the comparison above, a remarkable similarity appears between the latitudinal brightness measured by V1 in 1980 and the one measured by

UVIS 33 years later (Figure 6). In addition, the same trend is also observed by STIS on September 2, 2017 (e.g., Figure 6). The similarity obtained between the three distinct instruments measurements on different dates and different solar activity, after updating their calibration (Table 2), confirms that the observed latitudinal distribution is a permanent latitudinal pattern (bulge) of the Saturn Lyα disk brightness. The bulge is characterized by a brightness excess ~30% in the latitudinal range ~5-35N that is at least 12 σ above the brightness in the surrounding regions at the equator and ~60N. In addition, as shown in Figure 7, the distribution has no apparent dependency on the atmospheric longitude, a property of the Saturn Lyα disk brightness that confirms an earlier finding from Voyager UVS observations (e.g., Figure 6 in Yelle et al. 1986).

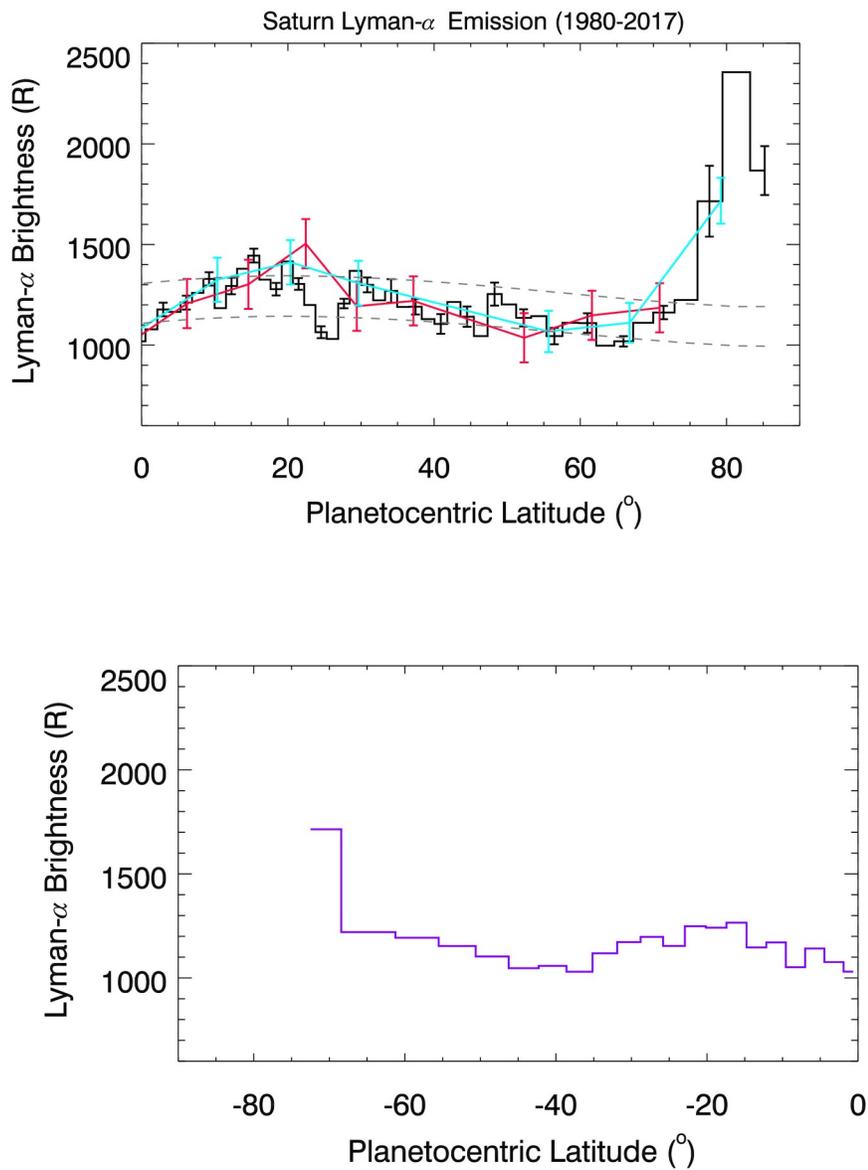

**Figure 6:** Saturn latitudinal Lyα brightness distribution. (Top) North summer hemisphere: (Aqua) STIS September 2, 2017. (Red) Voyager 1 UVS November 12, 1980 scaled by the solar flux ratio (~1.76) between the two dates and corrected for its calibration (~0.8). (Black) UVIS April 5, 2014 scaled by the corresponding solar flux ratio between the two dates (~1.41) and corrected for its calibration

(~1.74). The remarkable similarity obtained between the three distributions/instruments supports the existence of a permanent latitudinal brightness pattern that appears as a Lyα brightness excess (~30%) in the 5-35°N latitude range compared to equatorial and ~60N latitudes. The brightness peak around ~80N is not caused by any RT effect and is probably related to the auroral emission. (Dashed) RT model brightness for 2x and 3x the reference HI content. (Bottom) For the south summer hemisphere, we show UVIS September 5, 2007 scaled by the corresponding solar flux ratio between the two dates (~0.86) and corrected for its calibration (~1.78). The Lyα brightness distribution shows no clear latitudinal offset (O'Donoghue et al., 2019) compared to the northern hemisphere but is ~13% less pronounced. Unfortunately, we found no disk Lyα distribution from other instruments for the same south hemisphere and season.

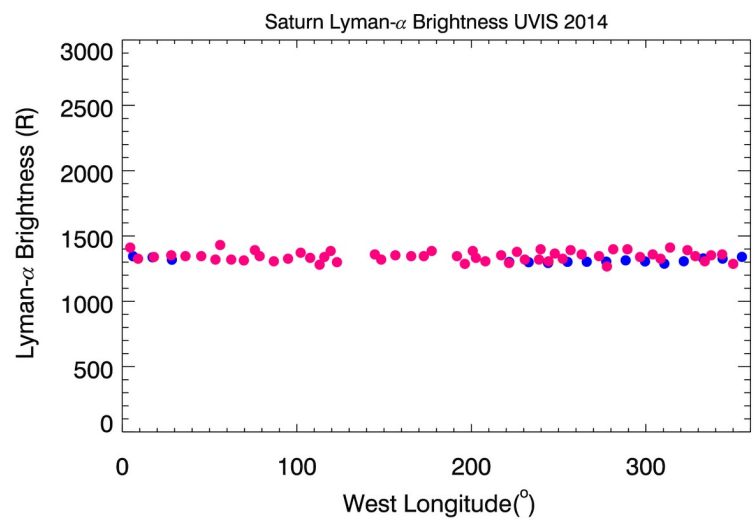

**Figure 7:** Saturn Lyα brightness distribution (blue) versus west longitude as observed by UVIS on April 5, 2014 (Table 1). For the covered longitude range, the distribution is fairly flat, much like the distribution derived from Voyager UVS

observation in 1980 (red), which strongly supports that the non-auroral Lyα emission shows no apparent variation with longitude.

To further test our diagnostic about the UVIS calibration, we now use a quite different approach based on the night-to-day (N/D) ratio of the Saturn brightness. This ratio is independent of the instrument calibration. Because both Saturn and IPH Lyα emissions are directly proportional to the solar source flux, the N/D ratio should be weakly dependent on the solar flux. In making this assumption, we implicitly require that the atmospheric reflectivity is evolving in the same way for the incident IPH and solar light sources at both low (STIS) and strong (V1) solar activity. For V1 (1980), we derived a ratio N/D~350 R/ 3300 R ~ 0.106 between night and dayside brightnesses (Broadfoot et al. 1981).

To check the validity of our assumption about the invariability of the N/D ratio, we searched the UVIS archive around maximum (similar to V1 conditions) and minimum (similar to HST/STIS 2017 observations) solar activity conditions. We could find quasi-simultaneous night (April 10, 2014) and day (April 5, 2014) observations of the Saturn disk close to the maximum solar activity at the same planetary latitudinal range with a ratio N/D~0.1065 (Tables 1 & 2). We also found quasi-simultaneous night (January 18, 2017) and day (January 16, 2017) observations of the Saturn disk close to minimum solar activity at a same planetary latitudinal range with a ratio N/D~0.1145, which is within 10% of the V1 value derived above. In the following, we use both values.

Thus, starting from N/D~0.106-0.114 measured by Voyager 1 in 1980 or UVIS in 2014 and 2017, we use the nightside measurement ~77.8 R made by UVIS on August 26, 2017 (e.g., Table 1) to derive a dayside brightness in the range ~ 730.5-679.5 R. When compared to the daytime measurement ~ 1230 R obtained by STIS on August 26, 2017 around the same latitudinal range (0-20°N), a correction factor in the range ~1.68-1.80 on the UVIS brightness is required to make the two measurements compatible. This compares rather well with the correction derived on the UVIS calibration at other dates (e.g., Table 2 and Figure 6).

As a final test of the UVIS calibration, we compare the interplanetary emission observed by UVIS on August 19, 1999 (when Cassini was near to the Earth orbit) to the sky background observed by HST/GHRS on June 4, 1994. Both instruments observed the sky background toward the so-called cross-wind direction (~90° from the interstellar medium upwind direction). After correction for the solar flux ratio between the two dates, we derive that a correction of ~1.6 is required on UVIS calibration in order to fit the GHRS IPH observation (Clarke et al. 1998).

For consistency, we also compare the interplanetary emission observed by GHRS on April 7, 1994 to the sky background observed by STIS on August

26, 2017. Both instruments observed the sky background toward the so-called upwind direction during solar minimum activity conditions (the interstellar medium upwind direction). After correction for the solar flux ratio between the two dates, we found no real difference (only a few %) between the two measurements (~960 R) (Clarke et al. 1998, and Appendix 2 for more details about the IPH line observed by STIS), which is a further indication of the consistency between the two instruments' calibration.

The conclusion on the correction ~1.75 required for the UVIS calibration when compared to HST spectrometers (GHRS & STIS) is robust because it is confirmed by different data-sets obtained on distinct dates in both similar and distinct solar activity. The attached error on the average correction factor for the UVIS calibration derived above is relatively small (statistical error less than 5%), yet we increase it to ~10% to account for potential systematic effects related to the assumptions made (line formation, reflectivity, etc.). Our finding harmonizes four decades of NASA space missions that observed the Saturn Lyα airglow over several solar cycles, with the discovery of a permanent latitudinal excess on its disk brightness that did not change between 1980 and 2017.

In the following, we use radiation transfer and atmospheric models to compare our brightness model to STIS/HST and archive data of different instruments, focusing on key properties of Saturn's disk brightness inferred in this section (latitudinal profile, line profiles, etc.).

## 4. Model comparison to observations

To analyze the HST/STIS Lyα data, we use 1D adding-doubling RT model that handles the scattering of photons with the constituents of Saturn's upper atmosphere that we describe with a photochemistry model. The RT model accounts for partial redistribution of photons during scattering with H, Rayleigh scattering of photons with $H_2$, and photoabsorption by hydrocarbon and water molecules (Ben-Jaffel et al., 2007). At the top of the atmosphere, Lyα solar flux and sky background Lyα emission at the orbital position of Saturn are required. For both the solar Lyα line (from Sun to Saturn) and for the Saturn Lyα line (from Saturn to Earth), we include the imprint of the interplanetary hydrogen absorption on the line profile (e.g., Section 3). For the Lyα sky background emission, we use in situ observations obtained by Cassini UVIS on August 27 and 31, 2017 (e.g., Table 1).

In addition, we use a one-dimensional (1D) photochemical model to describe the composition of Saturn's upper atmosphere as a function of altitude at specific latitudes during various time frames relevant to the Lyman alpha emission observations. The model utilizes the Caltech/JPL

KINETICS code (Allen et al., 1981; Yung et al., 1984) to solve the continuity equations for the chemical production, loss, and vertical transport of species. We develop several different models to test the sensitivity of the H abundance to different assumptions: (1) fixed-season 1D models at specific latitudes relevant to UVIS occultations described in Koskinen et al. (2015, 2016); (2) a time-variable seasonal model (e.g., similar to Moses & Greathouse, 2005) for 19 latitudes that considers neutral photochemistry only (adopting the neutral reaction list from Moses et al., 2018); (3) a time-variable seasonal model of the same 19 latitudes that considers coupled ion-neutral photochemistry (see Moses et al., 2022, for details of the ion-neutral chemistry model). The background atmospheric temperature structure for the latitudes considered in the seasonal model is derived from CIRS and UVIS occultation retrievals from observations late in the Cassini mission (Brown et al., 2020, 2022). In all models, the eddy diffusion coefficient profile is constrained by the methane retrievals from UVIS occultations (e.g., Koskinen et al., 2015; Brown et al., 2022) in the upper stratosphere and by ethane retrievals from CIRS data in the lower stratosphere (e.g., Fletcher et al., 2020). These models provide the vertical profiles of temperature and density of the constituents that contribute to the formation of the Lyα emission from the thermosphere of the planet—namely, H and $H_2$ for photon scattering and $CH_4$, $H_2O$, etc. for photoabsorption (e.g., Figure 8). To investigate the impact of a change in the gas mean mass, we also consider a few models with different He abundance (e.g., Figure 8).

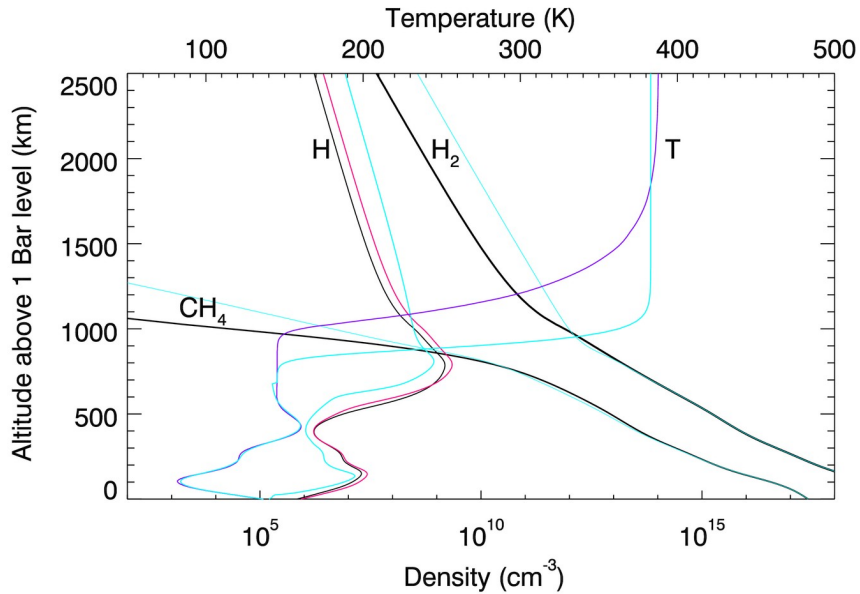

**Figure 8**: Example results from the photochemistry models of Saturn derived for 2°N planetocentric latitude. The 1D fixed-season models (red & black) adopt the thermal structure derived from a 2015 UVIS stellar occultation described in Koskinen et al. (2016), with an assumed deep helium abundance of 9.18%. The H atom abundance is sensitive to solar activity, and results are presented for solar minimum (black) and maximum (red) conditions. The increase of the solar flux from minimum to maximum activity enhances the HI abundance, which increases the Lyα albedo of the planet by ~10%. However, there is a degeneracy between temperatures and the assumed deep He abundance in the UVIS retrievals, as described more fully in Koskinen and Guerlet (2018). For our time-variable seasonal model at 2N at summer-solstice conditions (aqua), we adopt a thermal structure for this occultation (Brown et al., 2022) that assumes an updated 11% deep helium abundance, such that the resulting lower-thermospheric

temperatures must be increased in the "gap" region between the CIRS and UVIS data to account for the larger mean molecular weight of the gas and to maintain the $H_2$ density at higher altitudes consistent with the UVIS retrievals. The resulting temperature profile is steeper than that derived from Koskinen et al. (2016) in our fixed-season models, and the thermosphere extends deeper, which explains its much larger abundance of HI above the homopause level (above ~1000 km).

Ideally, one would solve the radiation transfer (RT) of the Lyα emission line using an oblate spheroid model for Saturn that takes into account the latitudinal variation of the altitude profile of the temperature and species densities as derived from the occultation observations. However, because the thickness of the H layer is very small compared to the radius of the planet, one could approximate the problem by taking several 1D plane-parallel RT calculations at specific latitudes and build the final brightness by interpolating between those single models. Yet this approximation has two limitations: first, it breaks down near the planetary limb because a single line of sight could cross different layers with distinct altitude scaling properties (gravity, temperature, etc.) versus latitudes; and second, fast latitude variation of the atmospheric parameters could be missed even if the H layer is thin.

Because our approach is driven by observations, we will investigate the relevance of any of those assumptions to our study case. For the nightside, we take the same dayside model because the chemical loss time scale for H is many years through the upper atmosphere, and although the production time scale is shorter at some altitudes – down to 90 hours at the µbar level – that is longer than a Saturn day. In addition, diurnally varying ions (of which $H_3^+$ is most important) do not significantly affect the local H abundance over the course of a day, although we find that ion chemistry itself increases the overall H abundance.

Practically, to compare the RT model results to the Lyα data, we generate several photochemistry models (fixed season) in the latitude range covered by the HST observations, namely at planetocentric latitudes (12S, 6S, 2N, 23N, 32N, 51N, and 70N). For each latitude, we calculate photochemistry models using solar XUV spectra for both minimum and maximum solar activity conditions (Woods & Rottman, 2002). For the time of the HST/CASSINI joint observations, the solar cycle was close to minimum activity. For the latitudes listed above, the corresponding total H column is [H]= (3.78,3.82,4.43,4.23,3.81,4.63) x$10^{16}$ cm$^{-2}$. Depending on the hydrocarbons homopause altitude, the HI content that contributes to the Lyα photon scattering is usually smaller.

For each atmospheric model, we ran the RT model and verified that Rayleigh scattering by $H_2$ has a negligible contribution. This result can be explained by the relatively small optical thickness of the $H_2$ layer available for Rayleigh scattering above the homopause level.

Because resonant scattering by atomic H appears as the dominant process for the Saturn Lyα line formation, it seems reasonable to investigate the possibility of a different atomic H column from that predicted by photochemistry models. Such thermal or non-thermal H content enhancement could result from many sources/processes such as auroral H production and global transport, precipitation of water or heavy H-bearing species from the ring system or Enceladus, high-resolution ultraviolet cross sections and solar flux (e.g., Kim et al., 2014; Chadney et al., 2022) not considered in our current models, high eddy diffusion coefficients extending into the thermosphere, seasonal modulation versus latitudes, and variation of the species scale height and the temperature vertical profile as discussed below. To facilitate the comparison with the HST/STIS observations (e.g., Figures 2 & 3), we thus conducted a simple sensitivity analysis versus the total H column using a factor 1x, 2x, and 3x the reference value at each specific latitude.

Interestingly, models corresponding to the reference atmospheric model (for solar minimum activity in 2017) cannot reproduce the observed brightness or the line profiles at all latitudes (e.g., Figures 2 & 3). However, models with 2 to 3 times the reference H content provide a rather good fit to both brightness and line profiles (e.g., Figures 2 & 3 and Table 3). Our result confirms the finding of earlier studies of the need for enhanced HI content to explain the Saturn Lyα airglow (Ben-Jaffel et al. 1995). For instance, doubling HI content is required to fit the high resolution line profiles at latitudes around ~2S, 55N, and 66N. In contrast, in the range 5 to 35 N, three times the HI content is required to fit the corresponding line profiles (e.g., Figure 3). Thus, a planet-wide enhancement of the H content is clearly required in order to fit the HST/STIS observations over the northern hemisphere up to 60N. Based on the RT modelling used here, the rather small HI column variation and exospheric temperature modulation with latitude in the considered photochemistry models are quite far from producing a modulation of ~30% contrast in the Lyα brightness as observed by V1, UVIS and STIS (e.g., Figure 6).

The challenge is then to find the potential sources of atomic H that could enhance the column in the range required by the STIS and UVIS observations, particularly at the latitude range ~5-35°. There are seemingly few options to enhance the H content in the thermosphere: either an external source that produces atomic HI or a substantial increase of the HI scale height in the region above the homopause level (Ben-Jaffel et al., 2007). During the grand finale orbits of Cassini, substantial influx of water (0.4-13.7 $10^8$ cm$^{-2}$ s$^{-1}$), methane (7-25.7 $10^8$ cm$^{-2}$ s$^{-1}$), and other heavy species have been reported around the equatorial region, showing the complexity of the coupling between the rings and the Saturn ionosphere/upper atmosphere (Waite et al., 2018; Yelle et al., 2018; Serigano et al., 2020). Such species modify the atmospheric structure and composition at specific latitudes (Yelle et al., 2018; Moses et al., 2022) and, according to the photochemistry model used here, enhance the H content. Independently, we also have a good indication of water precipitation from Enceladus into the upper atmosphere of Saturn, with an average downward flux of ~1.5 $10^6$ cm$^{-2}$ s$^{-1}$ planet-wide (Moses et al., 2000; Moore et al., 2015). To test both scenarios, we added a downward flux of water and methane on top of a few of our reference atmospheric models. Interestingly, influx of water from Enceladus in the range ~ $10^6$ cm$^{-2}$ s$^{-1}$ has negligible impact on the H content and Lyα brightness. However, the influx of water (~$3.7 \times 10^8$ cm$^{-2}$ s$^{-1}$) and methane (~$1.7 \times 10^9$ cm$^{-2}$ s$^{-1}$) in the range detected during the Cassini grand finale (Waite et al., 2018; Serigano et al., 2022) enhances the H content by a factor as much as ~1.75 compared to the reference model, which is close to the factor (2) required by the RT analysis at equator and mid-latitudes (50-60N) but still falls short of the value (3x) required in the 5-35N band. The problem is that the H enhancement required by the fit to the HST data is planetwide, while heavy species influxes were confined to specific latitudes (O'Donoghue et al., 2019). In addition, recent photochemistry modeling of species influx on Saturn's atmosphere clearly shows that substantial influxes of $H_2O$ above ~$10^7$ cm$^{-2}$ s$^{-1}$ are not supported by occultation observations, which tends to support the supposition that the measurements recorded by INMS were likely caused by small dust particles hitting the spacecraft and/or instrument during atmospheric passage and vaporizing (Moses et al., 2022). Considered in that way, it is difficult to link the HI enhancement required to fit the Lyα observation to $H_2O$ influx from the rings.

A second possibility for enhancing the HI content is to increase its scale height in the region above the homopause where most of the Lyα photons are backscattered. For reference, the structure and composition of the thermosphere are based on connecting two regions—namely, the region at pressures larger than a few $10^{-3}$ mbars that is probed by the CIRS limb scans and the region at pressures below a few $10^{-5}$ mbars that is best probed by UVIS occultation observations (Guerlet et al., 2011; Koskinen et al., 2015). In between, there is a gap where the temperature gradient is poorly constrained (Koskinen et al., 2015; Brown et al., 2020). If one adds the uncertainty on the gas mean mass and scale height that are sensitive to the He abundance, the resultant species local scale height and abundance could also be affected.

In the following, we test the scenario of a change in the thermal structure of the gap region, taking into account ion chemistry and seasonal effects

(over several Saturn years) that might also enhance the H content versus latitude depending on Saturn's season. As expected, the addition of ion chemistry does indeed result in more H, especially in the northern (summer) hemisphere. In addition, we observe much more HI in the seasonal model above the homopause because the whole atmospheric scale height (mean molecular weight) increased due to a larger deep helium mixing ratio (11%) used compared to the fixed-season model (9.8%). For instance, the temperature profile starts increasing deeper in the atmosphere, providing hotter HI in the gap region (yet with the same final exospheric temperature; e.g., Figure 8). The net effect is that the new HI distributions reflect enough Lyα photons to fit the Lyα line profile near the equator and 55N latitudes but still cannot reproduce the excess in the Lyα brightness observed in the 10-30N latitude range (e.g., Figure 9). However, the solution to the temperature profile in the gap region as used in the seasonal model is not unique. One can, in theory, adjust the temperature profile to improve the H atom abundance in order to enhance the Lyα brightness, as long as the overall $H_2$ density still remains consistent with the UVIS occultation and the gas scale height is also consistent with the He abundance assumed deep in the atmosphere.

As discussed above, the Lyα observations could be used to help break the degeneracy attached to the retrieval of the temperature vertical profile in the gap region between a few µbars and ~0.01 µbar. To achieve that goal, it is likely that using a global approach that associates distinct data sets is required. For instance, using occultation data together with He 584A airglow helped retrieve the deep atmosphere He mixing ratio for Jupiter (~16±3%) and Saturn (~15±2.5%) using specific temperature profiles based on occultation observations. For Jupiter, the derived He mole fraction (0.136) is consistent with the Galileo probe in situ value (0.137), while for Saturn, the derived He mole fraction (~0.13 ± 0.02) is larger than previously reported, making the stratosphere warmer by ~149K (Ben-Jaffel & Abbes 2015). Similarly, CIRS limb scans and UVIS occultation measurement have been combined to investigate the thermal structure of the thermosphere and the deep atmosphere helium mixing ratio (~11%) of Saturn (Koskinen & Guerlet 2018). More than likely, the next step is to consider an iteration between the CIRS limb scans, UVIS occultation data, and Lyα observation that should better constrain the temperature vertical profile, a task that is beyond of the scope of this preliminary investigation.

Besides the impact of the vertical temperature profile, our first test of a seasonal model indicates a potential role of ion chemistry and seasonal effect with a substantial enhancement in the H content above the homopause level (altitude ~1000 km). It is probable that sophisticated, time-consuming simulations that include solar cycle variations, aurorally produced H, enhanced ring-vapor inflow, and thermospheric circulation would be required to understand definitely the latitudinal distribution of H atoms in Saturn's thermosphere.

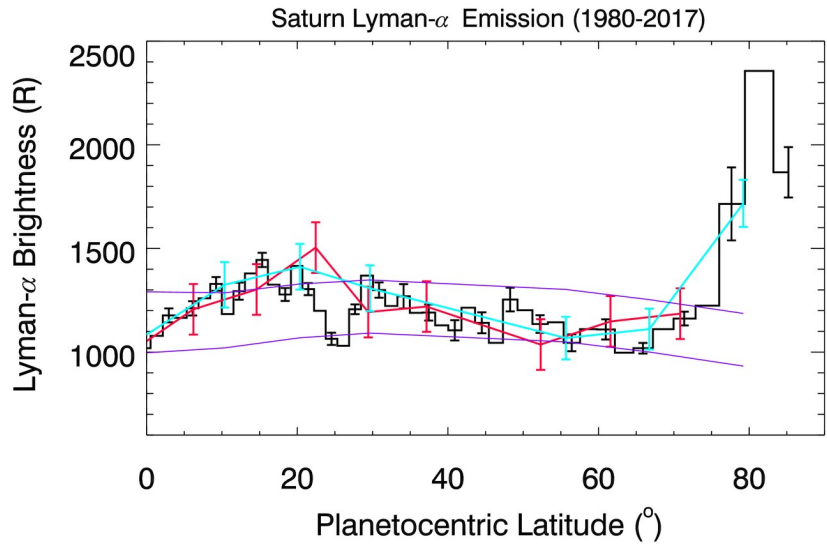

**Figure 9.** Saturn Lyα brightness in the north hemisphere (summer). Same as in Figure 6, except for the purple curves that represent the seasonal and ion chemistry model brightness for 1x and 2x the HI content. With the later model, we fit rather well the brightness levels near the equator and in the latitude range 50-65N. However, for the 5-35N latitude range, a larger HI content is required that could be produced by a steeper temperature gradient in the gap region between few μbars and ~10 nbars or by a thin layer (~$10^{12}$ cm$^{-2}$) of superthermal HI population in the region above ~1 nbar.

We also consider the impact of the presence of an arbitrary non-thermal H population that could be produced above pressure level ~1 nbar. The existence of this layer is justified by conditions (electromagnetic coupling with the rings, heavy species influx from the rings and Enceladus, etc) that favor the formation of hot atoms in the outermost region around the exosphere with a transition between a dense atmosphere and the collisionless region above it. For instance, the hot H column is negligible compared to the total H column of the atmosphere, yet the presence of hot atoms could affect the formation of the Lyα emission reflected by the atmosphere (Ben-Jaffel et al., 2007). The source of hot H atoms is defined by the strength of their non-thermal velocity. We are able to fit the Lyα brightness by invoking a latitudinal distribution of superthermal HI with a turbulent velocity ~18 km s$^{-1}$ (such as that produced from photo-dissociation of water; Crovisier, 1989); however, the origin of this population is arbitrary, and it is difficult to imagine that thermospheric winds would produce such a great level of turbulence (Sommeria et al. 1995). In addition, it is unclear how such population could survive thermalization by collision with the ambient gas. Kinetic ionospheric modeling is required to investigate some of these issues, a task that is also beyond the scope of this study.

At latitudes above ~60°, the influence of the auroral activity could be suspected at the origin of the brightness/width excess observed by HST/STIS. Indeed, the efficiency of the excitation of H/H$_2$ species by energetic particles impact that produce Lyα photons should decrease with latitudes from the main auroral north and south ovals (~+/-80°) down to latitudes ~+/-50°, respectively (Ben-Jaffel et al. 1995b; their Figure 2). We inspected the STIS spectrum for any weak auroral emission features but found no indication, probably because the detector thermal glow is the dominant source of contamination. The investigation of this potential process is beyond the scope of our study but should be addressed by future modeling of the auroral region based on the new HST/STIS observations obtained on September 2, 2017.

**Table 2:** Summary of Saturn Lyα brightness, solar flux variation, and inter-calibration factors during the period 1980-2017.

| UV Instrument | Date (day/year) | Orbital position (AU) | Angle/upwind (degrees) | Flux (@Sat) (avg +/-0.15A)*1.e9 ph/cm2/s | Calibration correction/reference instrument |
|---|---|---|---|---|---|
| IUE | 322/80 | 9.51 | 73. | 5.78 | 1.07/STIS |
| V1 | 317/80 | 9.52 | 73. | 6.093 | 0.8/STIS |
| V2 | 238/81 | 9.6 | 65. | 6.612 | 0.95/STIS |
| GHRS | 359/96 | 9.47 | 105. | 3.662 | 0.97/STIS |
| UVIS | 246/07 | 9.234 | 110 | 3.95 | 1.78/GHRS |
| UVIS | 133/13 | 9.83 | 39 | 4.056 | 1.73/STIS |
| UVIS | 154/13 | 9.83 | 41 | 3.973 | 1.75/STIS |
| UVIS | 95/14 | 9.9 | 43 | 4.892 | 1.75/STIS |
| UVIS | 238/17 | 10.06 | 7.0 | 3.053 | 1.7-1.8/Night/STIS |
| STIS | 238/17 | 10.06 | 7.0 | 3.053 | 1 |
| STIS | 245/17 | 10.06 | 7.0 | 3.463 | 1 |

## 5. Discussion & conclusions

We report HST/STIS Echelle high-resolution observations of the Saturn Lyα emission line obtained during a joint Cassini/HST campaign that occurred two weeks before the end of mission final plunge in 2017. From the STIS observations, we derive brightness and line profiles versus latitude that we use to compare to previous UV instruments (IUE, Voyager 1 & 2 UVS, HST/GHRS, UVIS) independently of any theoretical modeling. In this study, we searched the PDS UVIS archive to find UVIS brightness measurements that are obtained with light scattering conditions as close as possible to the ones observed by IUE, Voyager 1 & 2, HST/GHRS, and HST/STIS over the last four solar cycles (since 1980).

As shown in Table 2, our finding is that the Voyager 1 UVS sensitivity should be corrected by ~20% upward at the Lyα channels, the Voyager 2 UVS sensitivity should remain unchanged, and the Cassini/UVIS should be revised 75% downward if we take the HST instruments calibration as a reference. In addition, we found no need to correct the calibration of IUE and HST/GHRS compared to HST/STIS as the required correction is less than 5% (e.g., Table 2). With the revised calibrations, all detectors provide the right Lyα brightness level of Saturn at different epochs despite the evolving solar cycle.

The resulting corrections of the UVS brightness (~20% downward for V1 and ~4% for V2) are within the 30% uncertainty attached to the instruments' calibration (Ben-Jaffel & Holberg, 2016). In addition, Puyoo et al. (1997) discussed such a possibility of change in the UVS calibration and provided its impact on the far undisturbed interstellar medium (ISM) H density that should be revised from ~0.25 cm$^{-3}$ down to ~0.22 cm$^{-3}$ before any filtration in the outer heliosphere. In both cases, the derived ISM H density is consistent with the most recent radiation transfer investigation of the sky Lyα brightness distribution in the deep heliosphere as measured by Voyager 1 UVS (Katushkina et al. 2016).

This comparison also allowed the discovery of a permanent feature of the Saturn disk Lyα brightness that appears at all longitudes as a brightness excess of ~30% extending over the latitude range ~5-35N compared to the regions at equator and ~60N. This feature is confirmed by three distinct instruments between 1980 and 2017 in the north summer hemisphere of the planet. In contrast, in the southern summer hemisphere, the Lyα brightness shows less modulation. Interestingly, the Saturn latitudinal Lyα distribution was reported six years after the Voyager observation in 1980 and described as a distribution that shows no significant variation (Yelle et al., 1986). Finding the same feature observed by three different instruments was the key step in uncovering the Saturn bulge in the present study.

From the STIS observations, we derive brightness and line profile distributions versus latitude that we analyzed with a radiation transfer model combined with a latitude-dependent photochemistry model, itself constrained by Cassini/UVIS occultation observations. Our first finding is that the sophisticated photochemistry model does not produce enough atomic hydrogen to explain the Lyα brightness and line profile at all latitudes (planet-wide). By scaling the H distribution predicted by the photochemistry model, our finding is that 2 to 3 times the thermal H is required to reproduce the Lyα emission observed by HST/STIS at all latitudes, which calls into question the photochemistry model assumptions.

In an effort to explain the discrepancy, we considered the potential enhancement of the H content via planet-wide water influx from Enceladus on top of the atmospheric model (Moses et al., 2000; Moore et al., 2015). Based on the photochemistry model used here, such H enhancement is negligible and has no impact on the Lyα brightness. In the second step, we considered the possibility of H enhancement by influx of water and other species from the rings of the planet (Serigano et al., 2022). The same photochemistry model shows that the corresponding H content is important (factor ~1.75), yet such a large influx of water above $1.e7$ cm-2 s-1 seems inconsistent with UVIS occultation observations (Moses et al., 2022).

For latitudes above ~50°, auroral processes could potentially explain the bright and wide emission lines observed, but this requires confirmation by quantitative RT and auroral photochemistry modeling.

However, for latitudes below ~50° but different from those visited during the grand finale, we still need a global H enhancement. In addition to ion chemistry that is efficient to enhance the H content, here, we propose two additional potential scenarios that need confirmation in the future:
- The first scenario is related to increasing the species scale height in the region between a few µbars and ~10 nbars, a gap that is poorly constrained by ultraviolet occultations or infrared remote sensing observations. A tentative comparison with a photochemistry model using solar-cycle average solar flux values, ion chemistry, a deep atmosphere He mixing ratio of 11%, seasonal variations in geometry, and gaussian-shaped $H_2O$ influx as a function of latitude does provide the HI content required by the present Lyα analysis at several latitudes. More sophisticated simulations that include auroral production, ring-vapor inflow, thermospheric dynamics, a thermal structure that is consistent with the still-unconstrained He deep atmosphere mixing ratio, and ionospheric processes are needed to better understand Saturn's atomic H distribution.
- The second scenario is related to the production of superthermal HI population at high altitudes following the influx of materials from the rings or from Enceladus, or from turbulence produced from thermospheric winds. We have tested the potential effect of a thin hot HI layer that effectively enhances the planetary Lyα emission. However, it is not clear how this population is produced or how it could survive in the upper thermosphere.

Finally, our finding seem to confirm the main trend so far observed for exoplanets: their diversity. For instance, the Lyα bulge of Saturn is related to the influx of materials from the planet's rings, showing no significant longitudinal variation. In contrast, the Lyα bulge of Jupiter shows strong variation with the system III longitude of the planet, making the planetary magnetic anomaly the primary suspect at the origin of the process (Dessler et al., 1981). More generally, the HI content and the thermal structure of the extended atmosphere of most known exoplanets remain uncertain, with a degeneracy that existing models have difficulty resolving (Shaikhislamov et al. 2018, Ben-Jaffel et al., 2022). Our findings for the upper atmosphere of Saturn show the dramatic impact of materials influx from a ring system on the intrinsic composition/structure of the upper atmosphere of a giant exoplanet. In that frame, improving our understanding of the Saturn upper atmosphere (seasonal effects, thermospheric structure, etc.) is required in order to build the right diagnostic tools to properly interpret transit observation of a planetary system where a rings system is suspected.

## Appendix 1

First, we note that the calibration issue concerns the Lyα channels and not wavelengths shorter or longer than Lyα. Indeed, outside the Lyα channels, most instruments agree on the flux levels measured for stellar targets like Adara (epsilon CMa), a reference star that was used by KOS20 to show the consistency between IUE and UVIS. Here, we show the information that was missing: the signal observed by Voyager 2 UVS, for which the Lyα calibration was revised by a factor as large as 156% by Quemerais et al. (2013), is also consistent with the other instruments (Figure 10).

For reference, Lyα channels suffer significantly more impact from photons than the other channels do during an instrument's life, which influences the detector's local sensitivity over time. The use of a star like Adhara as a calibration lamp for the Lyα channels is not consistent with the fact that the stellar flux is nearly vanishing at those channels due to the absorption by the ISM hydrogen, despite the relatively smaller ISM H content along that line of sight. If any signal appears at those channels, it is the result of the line spread function of the low-resolution grating used and light scattering inside the instrument. At high resolution, HST/STIS observations of Adhara show almost no stellar photons at the Lyα channels (HST datasets: ocb6j0020 & ocb6j1020 for STIS/E140H and ocb6j0010, ocb6j1010, & ocb6j2010 for STIS/E140M).

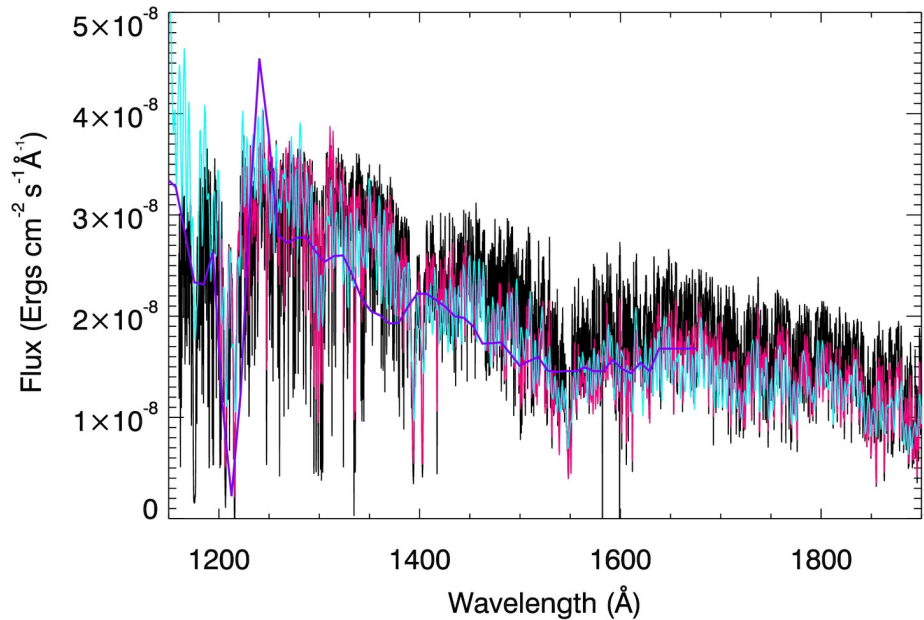

**Figure 10:** Comparison of the FUV spectrum of Adhara (ε CMa) observed by four NASA instruments: HST/STIS E140M (black), Voyager 2 UVS (purple), Cassini/UVIS (turquoise), and IUE (magenta). The flux levels recorded by the four instruments are consistent within ~20%, yet nothing could be identified on the Lyα channels that would require a specific investigation (see main text).

For those reasons, we believe that our technique, which is based on the Saturn Lyα brightness as a template lamp and uses HST/STIS as a reference instrument, is the most straightforward way to cross-calibrate past FUV instruments despite the different epochs.

## Appendix 2:

**STIS pipeline**: for each STIS Echelle spectral image obtained (3 for each HST visit of Saturn):

1-We read a flat-field corrected image as provided by the STSCI pipeline.

2- We obtain MAMA detector glow measured by STSCI over very long exposures during the same period (2017)[12].

3- We remove scaled 2017 glow image from the Saturn image. The scaling is applied based on the scattered light level measured in the fiducial bar spectral position (which blocks the source light and is only filled by photons scattered inside the detector).

4- We apply geometry distortion correction to result of step 3 (e.g., Figure 11; McGrath et al., 1998).

5- We use time-tag events to split the original exposure into two sub-exposures of equal exposure time. This is a new technique to derive the shape of the Earth's genuine geocoronal emission line at exactly the same spectral position on the MAMA detector. By subtracting one sub-exposure from the other, this technique affords the unique opportunity of removing any signal that is not varying over the exposure time (such as the interplanetary Lyα signal or the planetary emission) and only keeping the varying geocoronal emission line with a lower brightness level, yet with the exact emission line shape at exactly the same detector position.

---
[12] https://stars.stsci.edu/

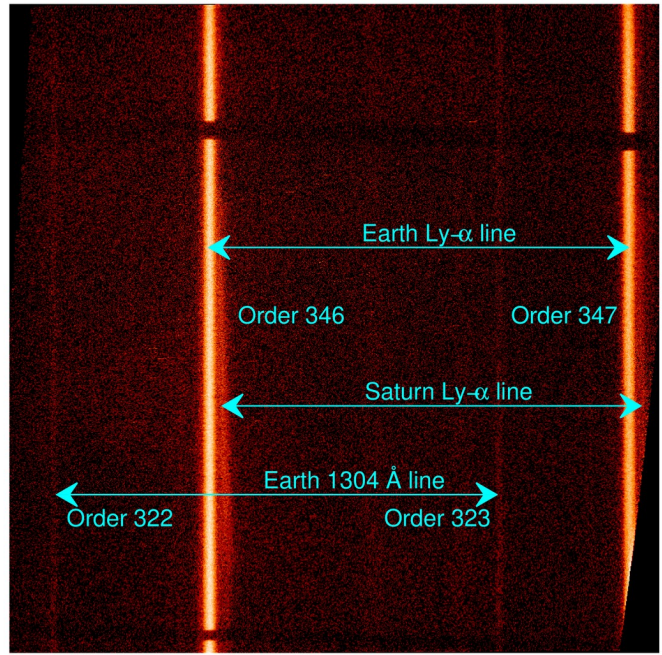

**Figure 11:** HST/STIS spectral image of Saturn Lyα obtained on August 26 2017. Corrected for geometry distortion, the image shows the Saturn and Earth emission lines for the two Echelle orders. Here, only order 346 is used.

6- We scale the genuine geocoronal line shape obtained in step 5 and remove from the initial observation to obtain the following results:

   6a- when using the sky background observation, the genuine geocoronal line derived from the same exposure allows one to properly derive the interplanetary emission line (between Earth and infinity), showing, for the first time, the self-absorption of the line by the Earth's geocorona atomic hydrogen (Figure 12). This effect is generally neglected and only rarely discussed in the literature. This step allows for reconstruction of the IPH emission line along that line of sight, which aids in its subtraction from the planetary signal.

6b-when using the Saturn observation (planet + sky background), our new technique allows one to properly derive the planetary emission after subtracting the genuine Lyα and interplanetary (between Earth and Saturn) using a least square fitting based on scaling the two line profiles obtained in the previous step. Usually, the IPH emission between Earth and Saturn is ~0.55 times the emission of the total IPH line, a fraction that seems weakly sensitive to the heliospheric angular position of Saturn (e.g., Table 2); see also Ben-Jaffel & Holberg 2016);

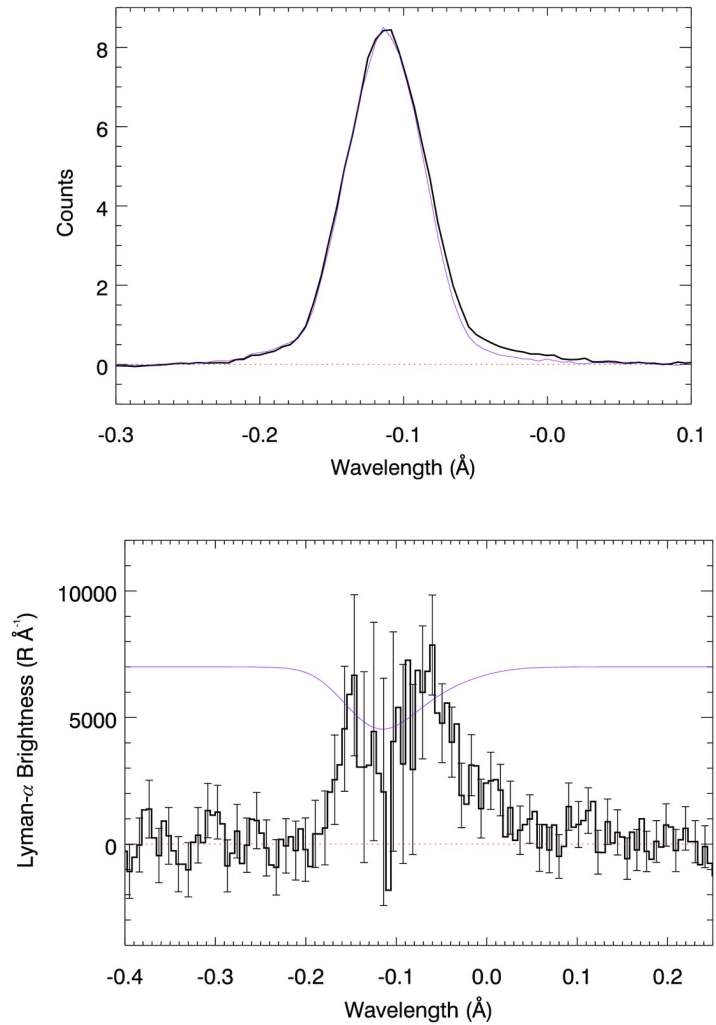

**Figure 12:** (Top): Sky Background Lyα line profile (black) observed on August 26, 2017 compared to the genuine geocoronal emission Lyα profile (purple) obtained by the new technique described in step 4 (heliospheric reference frame). The IPH Lyα emission appears as the difference between the two lines (red wing). (Bottom): The IPH Lyα line emission (black). We also show the self-absorption line profile by the geocoronal HI atoms that we scaled for clarity. This effect is usually neglected in the literature. The final IPH Lyα emission is 1060 R ± 130 R.

7-We apply step 4 particularly to the sky background observation obtained during each HST visit in order to derive a detector flat field correction along the STIS long slit, thereby ensuring that the sky signal should not vary spatially along the slit. We apply the same technique also to derive a flat-field along the UVIS slit.

8-We fulfill flux calibration using the STSCI official procedure described in the STIS data handbook (Section 5.4)[13] which we briefly describe in the following for the E140H grating when used with the STIS long slit 52"x0.5". Following the STIS data handbook, we define the surface brightness for an extended source as:

$$B_i = \frac{10^8 N_\lambda . h . c}{S_\lambda . A_{HST} . \lambda . f_{TDS} . f_T . W . m_s . disp}$$

where $N_\lambda$ is the count rate (ratio of total counts to the exposure time), h=6.626 $10^{-27}$ ergs/s is the Planck's constant, c=2.8879 $10^{10}$ cm s$^{-1}$ is the speed of light, $S_\lambda$ is the integrated system throughput as provided by the PHOTTAB file 15o1440ro_pht.fits (as defined in the STIS data file headers and that can be downloaded from the HST archive). In the expression above, $A_{HST}$=45238.93416 cm$^2$ is the area of the unobstructed HST mirror, λ is the wavelength (Å), $f_{TDS}$ =0.9 is the correction for time-dependent sensitivity of E140H, $f_T$=0.98 is the correction for temperature-dependent sensitivity for the same grating, m_s ~0.029 arcsec/pixel is the plate scale in the spatial direction, W =0.5 arcsec is the slit width, and disp~0.0054 Å /pixel is the Echelle grating dispersion.

---

[13] https://hst-docs.stsci.edu/stisdhb/chapter-5-stis-data-analysis/5-4-working-with-spectral-images

We have applied the above expression to the geometric corrected image derived in the previous step using order 346. Because of the geometric distortion, the Echelle order 347 also contains the Lyα line but is not used here because the spectral image is affected at the detector edge.

For consistency, we have tested the same procedure on the E140M observations that we obtained during the same HST visits and that are fully supported by the STSCI STIS pipeline calibration. Using the input parameters that are appropriate to that mode, we do obtain the same results as provided by the _x2d.fits files produced by the calstis pipeline.

The procedure is further tested using Echelle archive observations of stars (G191B2B, for example) obtained with slit mode 0.2x0.2" that is fully calibrated by the STSCI STIS pipeline to obtain consistent results.

9-To improve S/N, exposures obtained during a same HST visit are merged assuming that the planetary signal is not changing over a time period corresponding to 4.5 hours (3 HST orbits). We checked the individual exposures and noticed no time varying signal from the planet (within statistical noise attached to each exposure).

10-We fit the oblate shape of the planet at the time of observation using optical image of Saturn recorded during the same HST visit. This allows for allocating planetocentric latitudes/longitudes, incident solar light scattering angles with respect to local normals, and corresponding emission angles with respect to the line of sight to Observer (HST).

11-Doppler shifts due to the planetary spin are corrected for depending on the exact latitude/longitude position on the planetary disc as derived in the previous step.

12-Despite the improved S/N obtained at step 8-9, we found it necessary to bin spectra (Doppler shifted) each 50 pixels over the STIS slit. This corresponds to an average over a spatial region 1.2"x0.5" over the planetary disk.

13- Error bars include: statistical noise from photons counting, detector glow subtraction, Earth geocorona emission signal and interplanetary emission line at specific spectral ranges, and flat-field corrections provided either by the STSCI archives or derived in this study (step 6).

14- We do include the error due to the statistical uncertainty related to the position of the Lyα line center. Indeed, the final Lyα brightness depends on where the line center is placed in the observed spectrum.

**Appendix 3:**

In the following, we discuss few aspects of the UVIS calibration pipeline that may be useful in handling properly the software.

**1-** Cube Generator (CG) is the official pipeline to calibrate UVIS raw data[14]. It is available as a package that can be installed using the IDL software. A python version exists and can be accessed at https://github.com/Cassini-UVIS/pyuvis. One of the main problems related to the UVIS calibration is how to handle the response of the so-called "evil" pixels that behave in a way that is not fully understood. The UVIS team made great efforts to characterize their spatial distribution and time variation that led them to include additional time-dependent flat field correction to compensate for the deficient response of those pixels (UVIS Users Guide: Esposito et al., 2018).

For this project, we investigated the spatial and time evolution of evil pixels using sky background observations obtained between 1999 and 2017.

---

[14] https://pds-atmospheres.nmsu.edu/data_and_services/atmospheres_data/Cassini/CASSINIUVIS/UVIS-16/Cube%20Generator%20Software/

It appears from the different data sets that "evil" pixels modify the signal; some are repeatable, but others are not (such as those during 2008). Finding every single "evil" pixel over time is not trivial because of the additional statistical noise that makes the diagnostic difficult.

This means that CG and the extra flatfields collected from stellar observations are not enough to capture the sporadic behavior of each "evil" pixel. As far as we are interested in the integrated emission, we use long exposure IPH non-binned (1024 spectral pixels) observations where we merge the signal over the Lyα spectral band (including "evil" pixels), requiring that the final signal should not change over the spatial extent of the slit, beyond a linear trend across the limited spatial extent of the slit. This helps derive a pixel-to-pixel flat-field "evil" correction at the Lyα spectral band along the slit (spatial direction: 64 pixels) that we incorporate into the Cubegenerator software. In addition, we now propagate within Cube Generator the corresponding error bars that are not negligible. Working on the Lyα integrated signal helped us to obtain a decent S/N, which allowed for separation between statistical photon noise and sporadic "evil" pixels time-evolution.

We could find IPH low resolution (LR) data from 1999 to 2017, which allowed comparing the derived extra flat-fields over time. First, we confirm that CG provides a nice flat field outside the Lyα window (using the scattered light during each exposure). However within the Lyα window, there are changes as shown in Figure 13.

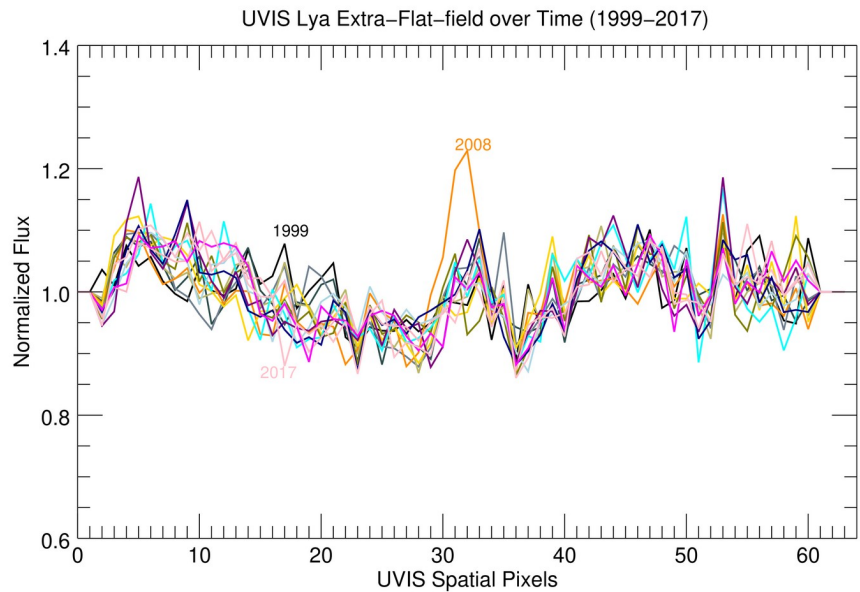

**Figure 13:** Pixel-to-pixel flat-field correction due to "evil" pixels affecting the Lyα spectral band over time (1999-2017).

As we can see, there is a same average trend for the spatial correction over time, except during 2008. We found that during 2008, a peak appears around UVIS spatial pixels 29-34 for all observations available for that period (a priori, it should not be star contamination because we tested exposures obtained from different regions of the sky).

But apart from that exception, we have the same trend that can be used to correct for the spatial distribution of the planetary Lyα brightness, which is related to the composition of the atmosphere.

One can also use UVIS high resolution (HR) observations of the sky background to derive an equivalent flat-field but we see no real difference because the IPH line is not resolved in any of the UVIS modes, in addition to limited time coverage in the HR mode.

**2-** Here, we assess the time evolution of the Lyα sensitivity as implemented in Cube Generator. In the spectral range outside Lyα, the UVIS team used repeated stellar observations to monitor the time evolution of the instrument sensitivity. The work done is described in detail in the UVIS Users Guide (Esposito et al., 2018). However, there are no known independent methods to perform a similar analysis at Lyα (Greg Holsclaw, private communication). To address that issue, we use the IPH dataset shown in Figure 13 to extract the calibration matrix provided by the Cube_Generator software over the time period between 1999 and 2017. Taking the UVIS 1999 sensitivity as a reference, we could derive the time evolution of the instrument Lyα response. As shown in Figure 14, the Lyα sensitivity implemented in CG declines by ~30% between 1999 and 2010 and remains unchanged up to the end of the Cassini mission. To avoid confusion, we stress that the calibration correction (~1.7) provided in our study is independent and should be applied on top of the sensitivity function implemented in the UVIS official pipeline.

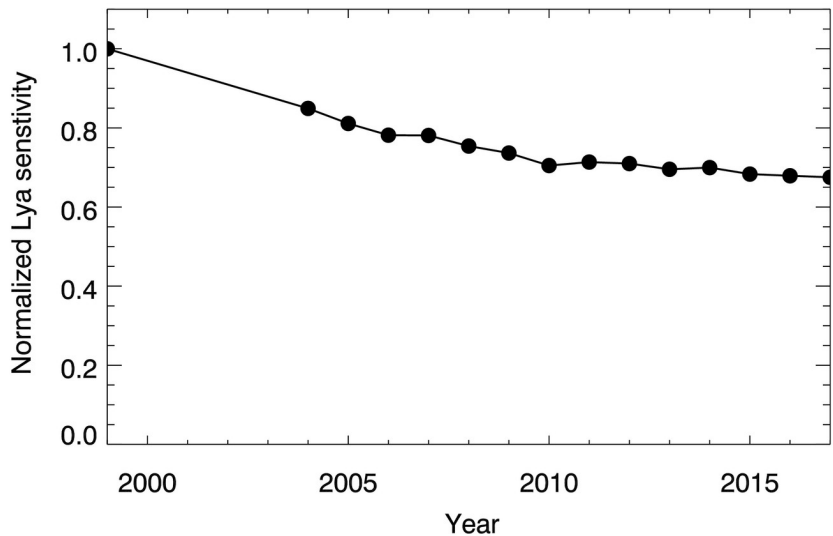

**Figure 14:** Yearly evolution of the UVIS Lyα relative sensitivity as implemented in Cube Generator, the UVIS official calibration pipeline. We show the sensitivity averaged over the Lyα spectral band relative to the 1999 sensitivity that we use as a reference (1999-2017).

**3-** By default, the UVIS calibration software removes a detector background of $4\times10^{-4}$ counts/second/pixel. Besides that noise level, we process the emission line by subtracting an average background from the adjacent red wing (pixels 95:110) and blue wing (pixels 150:165) of the line profile, and finally get the total brightness summing over pixels 114-145 that cover the whole line. We tested other spectral windows for both the adjacent background and the emission line and found no significant impact on our final results. This treatment is confirmed by our RT modeling that shows no significant contribution from flat intrinsic emissions such as by Rayleigh scattering at Lyα (e.g., Section 4).

## Acknowledgments:

We are very grateful to the staff at the Space Telescope Science Institute (STScI), and in particular to William Januszewski, Tony Sohn, and Shelly Meyett, for their careful work in the scheduling of our observations and for various instrument insights. We also acknowledge useful discussions regarding the UVIS calibration with T. Koskinen who also provided calibrated data for the 2017 UVIS observations. This helped us validate our Cube Generator software version. LBJ acknowledges support from CNRS and Centre National des Etudes Spatiales (CNES) in France under project PACES. J.M. and G.B. acknowledge support from STScI through Hubble Space Telescope grant GO-14931.007. J.M. gratefully acknowledges support from NASA Solar System Workings Program 80NSSC20K0462. These observations were also supported by STScI grants GO-16193-01 and GO-16195-01 to Boston University.

This work is based on observations with the NASA/ESA HST, obtained at the Space Telescope Science Institute (STScI) operated by AURA, Inc. Some of the data presented in this paper were obtained from the Mikulski Archive for Space Telescopes (MAST) at the Space Telescope Science

Institute. The specific observations analyzed can be accessed via https://doi.org/10.17909/cafj-3r46/